# Determination of the Riemann modulus and sheet resistivity by a six-point generalization of the van der Pauw method

Krzysztof Szymański, Kamil Łapiński, Jan L. Cieśliński, Artur Kobus*,
Piotr Zaleski, Maria Biernacka, Krystyna Perzyńska

*University of Białystok, Faculty of Physics, K. Ciołkowskiego Street 1L, 15-245 Białystok, Poland*

*\*Politechnika Białostocka, Wydział Budownictwa i Inżynierii Środowiska, ul. Wiejska 45E, 15-351 Białystok, Poland*

Abstract

Six point generalization of the van der Pauw method is presented. The method is applicable for two dimensional homogeneous systems with an isolated hole. A single measurement performed on the contacts located arbitrarily on the sample edge allows to determine the specific resistivity and a dimensionless parameter related to the hole, known as the Riemann modulus. The parameter is invariant under conformal mappings of the sample shape. The hole can be regarded as a high resistivity defect. Therefore the method can be applied for experimental determination of the sample inhomogeneity.

## 1. Introduction

Determination of direct relations between geometrical size or shape of the sample and electrical quantities is of particular importance as new measurement techniques may appear or the experimental accuracy can be increased. As an example, we indicate a particular class of capacitors based on the Thompson and Lampard theorem [1], in which cross-capacitance per unit length is independent of the size. Development of this class of capacitors leads to a new realization of the unit of capacitance and further the unit of resistance [2-4]. The second example is a famous van der Pauw method of measurement the resistivity of a two-dimensional, isotropic, homogeneous sample with four contacts located at its edge [5, 6]. The power of the method lies in its ability to measure the resistances with four contacts located at any point of the edge of a uniform sample. The result of the measurement is the ratio of a specific resistivity and a thickness. The method is widely used in laboratory measurement techniques. In fact, the van der Pauw method and the Thompson and Lampard theorem are closely related and follow from the solution of the Laplace equation in two-dimensional systems with well-defined boundary conditions [7].

In the original concept of van der Pauw, two four probe resistances are measured on the flat, homogeneous sample without isolated hole. A kind of nonlinear equation including the resistances, allows determination of an important parameter: ratio of a specific resistivity and the sample thickness. The van der Pauw method is widely used in laboratory measurement techniques. However, strict assumptions of the method - point contacts and the sample homogeneity - are difficult to obey in experimental practice. The resistivity measurement is a weighted averaging of local resistivities. Koon *et al* has introduced concept of a weighting function used for estimation of the resistance as a weighted averaging of local resistivities. A formalism to calculate the weighting function for van der Pauw samples was proposed and measured in [8]. Although the weighting function has to be calculated for arbitrary sample shape [9], it is important to note that the weighting function is conformally invariant. Many authors have demonstrated that sample inhomogeneity or defects results in incorrect value of the measured resistivity and proposed some correcting procedures [9,10,11,12]. To our best knowledge all of the proposed corrections depend on the specific sample shape and thus spoil the generality of the van der Pauw approach.

The van der Pauw-type measurements can be extended on doubly-connected samples (samples with a hole) [13]. Such samples are characterised by a dimensionless parameter known as the Riemann modulus [14]. Any two-dimensional sample with an isolated hole can be transformed by conformal mapping into an annulus (see [15], page 255) with inner radius $r_{inn}$ and outer radius $r_{out}$, and the Riemann modulus $\mu = r_{out}/r_{inn}$. The Riemann modulus is therefore an invariant of the class of two dimensionless samples with a hole. Since a sample with a hole is a natural extension of a homogeneous sample without an isolated hole (considered originally by van der Pauw) the Riemann modulus of the sample without hole is equal to infinity.

Recently it was shown [16] how to determine the sheet resistivity and Riemann modulus for doubly-connected sample of arbitrary shape. A pair of four probe resistances measured for various locations of probes fall inside a certain region which is confined by some boundaries. The essence of the method [16] was observation, that some extreme values of the resistances show well defined correlation. The shape of the correlation curve depend on both, the Riemann modulus and the sheet resistance [16]. The possibility of determination of the Riemann modulus is important for practical applications because it can be interpreted as a measure of the sample inhomogeneity. However practical realisation described in [16] is very inconvenient because one has to perform many changes of the positions of the contacts to get many measurements of four probe resistances. In this paper we propose a new, six point method. The contacts located at arbitrary positions on the sample edge allows in principle determination of the geometrical parameter and the specific resistivity in a single measurement.

## 2. Six point method

Let us consider conductive medium with four different point contacts located at points $\alpha\,\beta\,\gamma\,\delta$. A current $j_{\alpha\beta}$ enters the sample at the contact $\alpha$ and leaves at the contact $\beta$, while the potential $V_{\gamma\delta}$ is measured between contacts $\gamma$ and $\delta$. The resistance for contacts $\alpha\,\beta\,\gamma\,\delta$ is defined as $R_{\alpha\beta\gamma\delta} = V_{\gamma\delta}/j_{\alpha\beta}$. We further consider only ohmic media, e.g. currents are linear functions of the potentials. It was shown recently [13] that in a special case of the four contacts located at the edge of an annulus with outer radius $R$ and inner radius $r$ (shown in Fig. 1 *a*)

$$R_{\alpha\beta\gamma\delta} = \lambda \ln \frac{G(\gamma-\alpha,h)G(\delta-\beta,h)}{G(\gamma-\beta,h)G(\delta-\alpha,h)}, \qquad (1)$$

where $h$, is a geometric parameter defined in [17] and $\lambda = \rho/(\pi d)$ is proportional to the specific resistivity $\rho$ of a sample with thickness $d$. The function $G(\varphi,h)$ is defined as

$$G(\varphi, h) := \left|\sin\frac{\varphi}{2}\right| \prod_{n=1}^{\infty}\left(1 - \frac{\cos\varphi}{\cosh hn}\right). \qquad (2)$$

The geometric parameter $h$ is directly related to the so called Riemann modulus of a doubly connected region. Any doubly connected region can be conformally mapped into an annulus with outer radius $R$ and inner radius $r$. The ratio $\mu := R/r$ ($\mu > 1$), known as the Riemann modulus, is invariant with respect to conformal transformations. The parameter $h = 2\ln\mu$ for an annulus. The function $G$ is even and periodic:

$$G(-\varphi, h) = G(\varphi, h), \quad G(\varphi + 2\pi, h) = G(\varphi, h). \qquad (3)$$

Taking into account the evenness of $G$ we easily see that $R_{\alpha\beta\gamma\delta} = R_{\gamma\delta\alpha\beta}$, $R_{\alpha\beta\gamma\delta} + R_{\beta\delta\gamma\alpha} + R_{\alpha\delta\beta\gamma} = 0$ which is a particular case of the general reciprocity theorem [5].

Let us consider the same annulus with six contacts located at arbitrary angles of the same the edge, shown schematically in Fig. 1 $b$. In our convention $\alpha$ $\beta$ $\gamma$ $\delta$ indicate the positions of contacts in Fig. 1 $a$ while $\varphi_i$, $i=1, \ldots, 6$ are angles between the positions of the contacts in Fig. 1 $b$. There are only five independent differences $\varphi_i$, since $\varphi_1 + \varphi_2 + \varphi_3 + \varphi_4 + \varphi_5 + \varphi_6 = 2\pi$. By inspection of eq. (1), linearity of the system and above mentioned reciprocity relations, one can easily check that contacts located at the vertices of the inscribed hexagon in Fig. 1 $b$, results in nine four probe resistances shown schematically in Fig. 1 $c$. Six of them are obtained by cyclic permutations and are abbreviated by $r_i$, $i=1, \ldots, 6$ while three others are abbreviated by $r_{i,j}$, were pair of indices $(i,j) = (1,4), (2,5), (3,6)\ldots(6,3)$ correspond to location of contacts at opposite edges of the hexagon. It is clear that one consequence of our definition is $r_{i,j} = r_{j,i}$. Nine resistances shown in Fig. 1 $c$ depend on the geometrical parameter $h$, resistivity $\lambda$ and five independent differences $\varphi_i$. Thus we obtain the following overdetermined system of 9 equations for 7 unknowns:

$$r_i = \lambda \ln \frac{G(\varphi_i + \varphi_{i-1}, h) G(\varphi_i + \varphi_{i+1}, h)}{G(\varphi_i, h) G(\varphi_{i-1} + \varphi_i + \varphi_{i+1}, h)},$$

$$r_{i,i+3} = \lambda \ln \frac{G(\varphi_i + \varphi_{i-1} + \varphi_{i-2}, h) G(\varphi_i + \varphi_{i+1} + \varphi_{i+2}, h)}{G(\varphi_{i-1} + \varphi_{i-2}, h) G(\varphi_{i+1} + \varphi_{i+2}, h)}, \qquad i=1,2,\cdots., \qquad (4)$$

where we use $\varphi_{i+6} = \varphi_i$ in order to define $\varphi_i$ for indices outside the range $1, \ldots, 6$. The same rule can be applied to $r_i$ and $r_{i,j}$ when needed. Taking into account $\varphi_1 + \varphi_2 + \varphi_3 + \varphi_4 + \varphi_5 + \varphi_6 = 2\pi$ and (3) we have

$$G(\varphi_{i+3} + \varphi_{i+2} + \varphi_{i+1}, h) = G(\varphi_i + \varphi_{i-1} + \varphi_{i-2}, h)$$
$$G(\varphi_{i+3} + \varphi_{i+4} + \varphi_{i+5}, h) = G(\varphi_i + \varphi_{i+1} + \varphi_{i+2}, h). \qquad (5)$$

Hence $r_{i,i+3} = r_{i+3,i}$ which perfectly agrees with the above physical argument.

The currents and potentials in (1) are invariant under any conformal transformation. Therefore eqs (4) are valid for a sample of any shape with an isolated hole. Having measured resistances $r_i$ for arbitrary position of the contacts, the parameters $h$, $\lambda$ and $\varphi_i$, can be obtained by solving a set of nonlinear algebraic equations (4). Since the set (4) is overestimated, the convenient method of solution is minimization of

$$\chi^2 = \sum_{i=1}^{6}\left(r_i - r_i^{exp}\right)^2 + \sum_{j=1}^{3}\left(r_{j,j+3} - r_{j,j+3}^{exp}\right)^2 \qquad (6)$$

with respect of 7 independent parameters: $h$, $\lambda$ and six angles $\varphi_i$ constrained by $\varphi_1+\varphi_2+\varphi_3+\varphi_4+\varphi_5+\varphi_6=2\pi$. The symbols $r_i^{exp}$ and $r_{j,j+3}^{exp}$ were used for abbreviation of the experimentally measured four-probe resistances.

It is worth mentioning that in the analogous five-point method, one get five independent resistances and six unknowns. Therefore the system is underdetermined and one cannot determine all required quantities. Thus the presented six point method, although resulting in overdetermined system of equations (4), is an optimal one.

## 3. Experimental results

Our samples were prepared from commercially available, technical quality, brass foil of the thickness (0.271±0.003)mm the form of disc with radius (146.1±0.2) mm and annulus with outer radius $R$=124.5±0.5 mm and inner radius $r$=74.3±0.3 mm. The geometrical parameter for the annulus depends on the ratio of radii [13] and in our case $h=2\ln(R/r)=1.03\pm0.01$. Resistivity to thickness ratio was determined by the standard van der Pauw method (for the disc) and also by the envelope method (proposed in [16] for the annulus) giving the same value $\lambda$=(0.3489±0.0003) m$\Omega$.

The six point method was verified on the annulus by comparing values of $\varphi_i$ obtained by minimization of (6) and experimental positions of contacts located at the outer boundary as shown schematically in Fig. 1 *b*. In order to simplify the comparison, positions of the contacts were parametrized by one parameter $\varphi$, see Fig. 1 *d*. The complete set of experimental results is presented in Fig. 2. We see a very good agreement of experimental data with theoretical dependencies given by Eq. (4). Note that for the specific configuration of contacts shown in Fig. 1d our theory predicts: $r_1=r_5$ and $r_2=r_3=r_4$. Moreover, $r_1$ and $r_5$ should vanish for $\varphi=2\pi/5$ (i.e., $\varphi_6$=0). All these properties can be observed at Fig. 2, as well. There are clear differences between results shown in Fig. 2 *a* and 4 *b*. Therefore the presented method is sensitive enough for detection of a hole. In the case of the configuration from Fig. 1 *d*, where positions of the contacts are parametrized by a single value $\varphi$, we expect that all five values $\varphi_i$ (i=1,2,…5) minimizing (6) are equal to $\varphi$. The sets of solutions for different values of $\varphi$ are presented in Fig. 3. We see that for $\varphi$ <0.3 the agreement with the theory is far from being perfect (it corresponds to configurations where all 6 contacts are located in near each other, in a quarter of the circle). The highest precision is obtained when at least 5 contacts are approximately equally spaced on the edge of the annulus ($\varphi=2\pi/6$). The obtained value of $h$ agrees (for $\varphi$>0.3) with geometrical value $h$ determined from ratio of the radiuses.

The six point method was also applied for the sample with an irregular shape. Nine resistances were measured for arbitrary positions of the contacts (shown on ordinate in Fig. 3 *c*) and geometrical parameter $h$ and five unknown angles were obtained from (5). Next the predicted resistances obtained from were drawn on abscissa of Fig. 3 *c* showing perfect agreement.

## 4. Discussion

One of the most obvious application of the presented method is the determination of the parameter of geometrical type known as the Riemann modulus by measurement of the resistances between contacts located at the sample edge. One may expect not too many applications for determination of the Riemann modulus of the sample with a hole. Nevertheless there is at least one important case. Small hole can be regarded as a high resitivity defect in an otherwise homogeneous sample. Therefore, the hole is mimicking a class of defects. A small hole can be considered as a small deviation from the homogeneity and the procedure of measuring the Riemann modulus can be used for quantitative characteristic of the sample quality. Perfectly homogeneous sample without a hole should yield the Riemann modulus equal to infinity.

The important question is about sensitivity of the method on the presence of small hole. We have performed initial experiments and we have observed clear dependence of the Riemann modulus on the hole size when the hole was relatively large. However in case of small hole located at the sample center, the experimental observation are disturbed, most probably by sample inhomogeneity, by distribution of the thickness and, as a consequence distribution of the lambda parameter, e.g. ratio of specific resistivity and the sample thickness. The experiments are in progress.

The presented formalism is applicable to two dimensional systems with an isolated hole. However it does not mean that only thin, flat samples can be measured. A samples made of thin layer forming a cylindrical [13] or conical surface can be measured by positioning six contacts on one edge as shown schematically in Fig. 4 *a*, *b*, yielding the Riemann modulus and a sheet resistivity.

Also a class of samples of arbitrary thickness (Fig. 4 *c-f*) can also be considered as two-dimensional systems. The six point method can be applied by using the knife edge contacts shown schematically in Fig. 4 *c*. Therefore large class of elements – thick cylinders and thick cylinders with cylindrical hole (symmetric or eccentric, see Fig. 4 *e*, *f*). Also the thick cuboid (with a hole or without a hole) shown in Fig. 4 *d* can be measured by knife-shaped contacts. It is worth mention, that one of the possible applications is the testing of the large and thick steel sheets produced industrially by rolling. These large and heavy objects can be tested when stored on the stacks where only the sheet edges are easily accessible. Another obvious application areas is an option of six point measurements in scientific laboratory equipment using the standard van der Pauw method. In this way sample quality could be tested simultaneously.

There are nondestructive methods of investigation of local resistance by four probes located far from the sample edge, see [17] and references cited therein. In contrast to these measurements, proposed six point method is global, e.g. an inhomogeneity degree depend on the whole sample. In other words, in a single measurement the hole can be detected irrespectively of their location. Therefore presented method is complementary and not competitive to the already mentioned technique. The advantage of the six point method is a single, well defined quantity, the parameter *h* in (4) considered as the inhomogeneity degree.

5. **Summary**

We have presented and verified experimentally an extension of the van der Pauw method. The extension is valid for two dimensional, flat, homogeneous samples with an isolated, single hole. In the method six contacts are located at arbitrary positions of the sample edge and by measurement of resistances between the contacts one is able to determine the geometrical parameter known as the Riemann modulus and the resistivity to thickness ratio. The geometrical parameter is invariant under conformal transformation.

The method was verified experimentally on a sample with shape of annulus where positions of contacts were used for comparison between the measured and predicted values of the geometrical parameter. Full agreement was achieved. Results presented in Fig. 2 *a* and 2 *b* clearly show that the method is sensitive to the presence of an isolated hole.

**Figures**

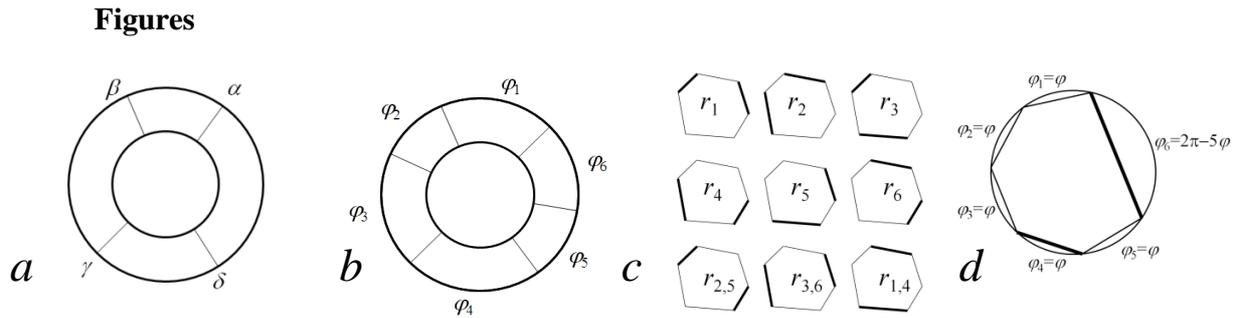

Fig. 1. *a*) An annulus with four contacts at arbitrary positions $\alpha$ $\beta$ $\gamma$ $\delta$ on the same edge. *b*) An annulus with six contacts. Symbols $\varphi_i$ denote angles between contacts (not the angular positions). *c*) Schematic positions of contacts for nine independent resistances. Each pair of current contacts or voltage contacts is located at the ends of the thick bars. *d*) Schematic positions of six contacts parameterized by one parameter $\varphi$. As an example, thick bars show position of contacts for measurement of resistance $r_5$.

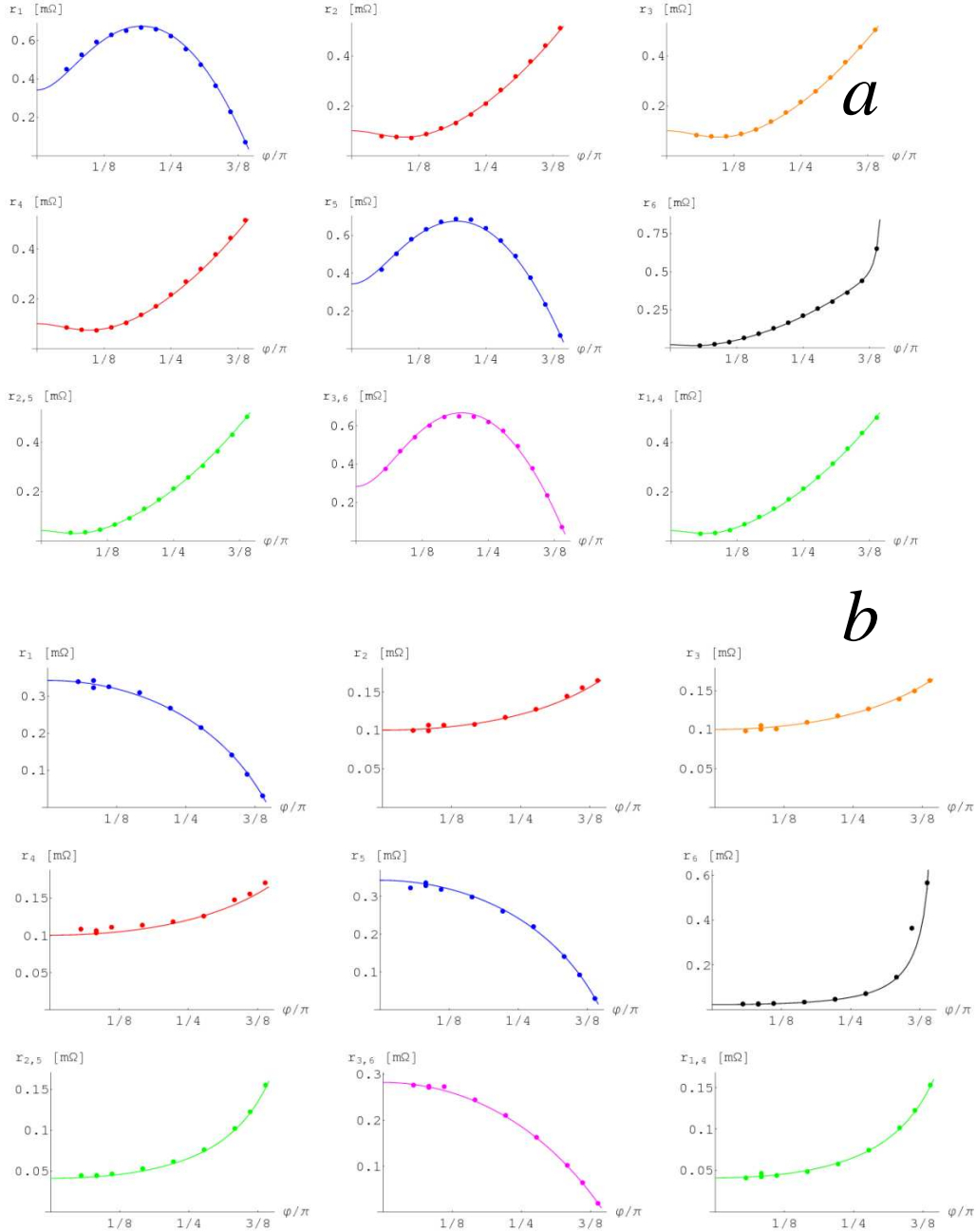

Fig. 2. Resistances $r_i$, $r_{i,j}$ (defined in Fig. 1 *c*) measured (in the case shown in Fig. 1 *d*) as functions of $\varphi$ on a) annulus and b) disk. Solid lines show theoretical curves given by Eq. (4), where $\lambda=0.3489$, $\varphi_1 = \varphi_2 = \varphi_3 = \varphi_4 = \varphi_5 = \varphi$ and $\varphi_6 = 2\pi - 5\varphi$, and parameter $h=1.03$ in a) and $h=\infty$ in *b*).

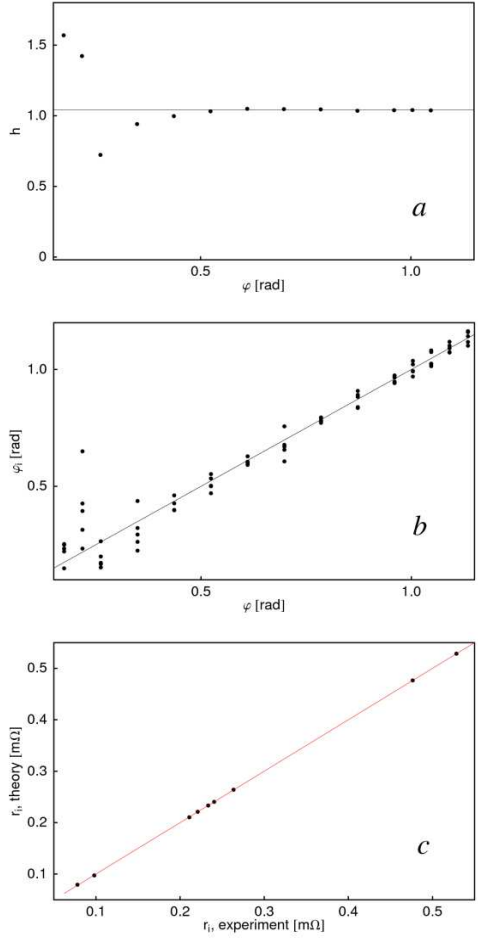

Fig. 3. *a*) Geometrical parameter $h$ obtained as a solution of nonlinear system (4) by the minimization of (6) for six contacts located at the edge of the annulus under experimental conditions $\varphi_1=\varphi_2=\varphi_3=\varphi_4=\varphi_5=\varphi$, $\varphi_6=2\pi-5\varphi$. *b*) Angles $\varphi_1$, $\varphi_2$, $\varphi_3$, $\varphi_4$, $\varphi_5$ between contacts estimated by minimization of (6). For each value of $\varphi$ (ordinate) there are five estimated values (abscissa) shown by dots (some of them overlap). *c*) Comparison of resistances $r_i$, $r_{i,j}$ measured on irregular sample with theoretical predictions, where $h$, $\lambda$ and five angles are obtained from minimization of (6). Solid lines in *b*) and *c*) are guides to eye corresponding to the ideal agreement between experimental values and estimation based on the minimization of (6).

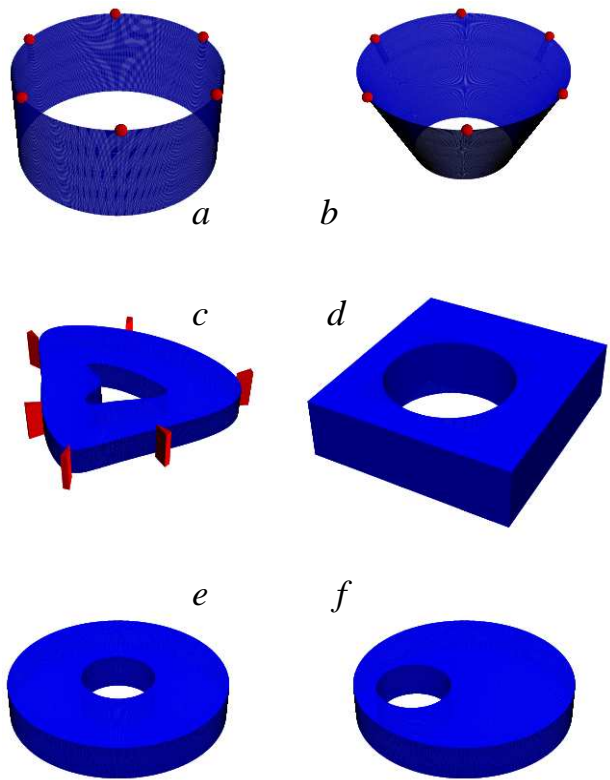

Fig. 4. Example of a sheet forming a sample of *a*) cylindrical and *b*) conical shape with six contacts located at one sample edge (red points). *c*) An example of thick sample of arbitrary shape with a hole, equivalent to two dimensional system. Six linear contacts in the form of knife edges are shown by red color. *d*) *e*) *f*) some other thick samples equivalent to two dimensional systems with linear contacts used during the six point measurements.

**References**

[1] A. M. Thompson and D. G. Lampard 1956, A new theorem in electrostatics and its application to calculable standards of capacitance, *Nature* **177** 888
[2] W. K. Clothier 1965, A calculable standard of capacitance, *Metrologia* **1** 36-55
[3] A. M. Thompson 1968, An absolute determination of resistance based on a calculable standard of capacitance, *Metrologia* **4** (1) 002 1-7
[4] H. Bachmair 2009, Determination of the unit of resistance and the von Klitzing constant $R_K$ based on a calculable capacitor, *The European Physical Journal Special Topics* **172** 257-266
[5] L. J. van der Pauw 1958, A method of measuring the resistivity and Hall effect of discs of arbitrary shape, *Philips Research Reports* **13** (1) 1-9
[6] L. J. van der Pauw 1958, A method of measuring specific resistivity and Hall coefficient on lamellae of arbitrary shape, *Philips Technical Review* **20** 220-224

[7] J. J. Mareš, P. Hubík and J. Krištofik 2012, Application of the electrostatic Thompson–Lampard theorem to resistivity measurements, *Meas. Sci. Technol.* **23** (4) 045004

[8] D. W. Koon and C. J. Knickerbocker 1992, What do you measure when you measure resistivity?, *Rev. Sci. Instrum.* **63** 207-210

[9] D. W. Koon, F. Wang, D. H. Petersen, and O. Hansen 2013, Sensitivity of resistive and Hall measurements to local inhomogeneities, *J. of Appl. Phys.* **114** 163710

[10] O. Bierwagen, T. Ive, C. G. Van de Walle and J. S. Speck 2008, Causes of incorrect carrier-type identification in van der Pauw–Hall measurements, *Appl. Phys. Lett.* **93** 242108

[11] J. Náhlík, I. Kašpárková and P. Fitl 2011, Study of quantitative influence of sample defects on measurements of resistivity of thin films using van der Pauw method, *Measurement* **44** 1968-1979

[12] F. Wang, D. H. Petersen, M. Hansen, T. R. Henriksen, P. Bøggild, and O. Hansen 2010, Sensitivity study of micro four-point probe measurements on small samples, *J. Vac. Sci. Technol.* **B 28** C1C34-C1C40

[13] K. Szymański, J. L. Cieśliński and K. Łapiński 2013, Van der Pauw method on a sample with an isolated hole, *Phys. Lett.* **A 377** 651-654

[14] Z. Nehari 1952, *Conformal mapping* (New York: Mc Graw-Hill)

[15] L. V. Ahlfors 1979, *Complex analysis* 3rd edn (New York: Mc Graw-Hill)

[16] K. Szymański, K. Łapiński and J. L. Cieśliński 2015, Determination of the Riemann modulus and sheet resistance of a sample with a hole by the van der Pauw method, *Measurement Science and Technology*, in press

[17] N. Bowler, 2011, Four-point potential drop measurements for materials characterizatio, *Measurement Science and Technology* **22** 012001